# Metasurface tessellation for reconfigurable electromagnetic scattering

**XIANGDONG KONG,[1] CHUANNING NIU,[1] JIA ZHAO,[1] AND ZUOJIA WANG[1,2]\***

[1]*School of Information Science and Engineering, Shandong University, Qingdao 266237, China*
[2]*Key Laboratory of Advanced Micro/Nano Electronic Devices & Smart Systems of Zhejiang, The Electromagnetics Academy at Zhejiang University, Zhejiang University, Hangzhou 310027, China*

*\*zuojiawang@zju.edu.cn*

**Abstract:** Metasurfaces have attracted significant research interest owing to their unprecedented control over the spatial distributions of electromagnetic fields. Herein we propose the concept of metasurface tessellation to achieve reconfigurable scattering functions. Square meta-tiles, composed of identical structures, are arranged to fill a surface. The electromagnetic scattering of the tiled surface is determined by the orientation distribution of the meta-tiles. We present three typical cases of meta-tiles consisting of binary elements to realize several distinct scattering patterns. This study provides an alternative method to build reconfigurable and multi-functional metasurface devices without external stimuli and complicated fabrication.

## 1. Introduction

Metasurfaces are composed of subwavelength units arranged according to pre-designed patterns, which have been introduced in the area of optics technology to manipulate the wavefront of light [1, 2]. By tuning the unit geometries, the electromagnetic (EM) responses of metasurfaces can be tailored [3, 4]. The unique physical properties and flexible EM controllability have attracted enormous interest because of their ability to manipulate the microwave or visible light propagation [5]. Metasurfaces have been applied in the microwave and optical systems to achieve various functions, such as invisibility cloaks [6, 7], holograms [8, 9], high-performance antennas [10, 11], perfect absorption [12-14], focusing lens [15, 16], frequency selection[17], chiroptical responses [18-20], and many other devices and systems.

The conventional metasurface design process, which consists of model design, parameter sweeping, and optimization, is complicated because a large number of elements with distinct geometries are needed. Some efficient design methods focus on utilizing optimization algorithms such as genetic algorithms [21-23] and deep learning [24]. Another inconvenience of the traditional metasurface is that once manufactured, every single device performs only one function for which it was designed earlier. A solution to this problem is to make a reconfigurable metasurface. Relevant approaches include rotating the metasurface around an axis by a motor [25], or tuning the orientation of the metasurfaces using a micro-electro-mechanical system [26]. Both methods only change the position or orientation of the metasurface, but never transform its structure or its parameters. In 2014, based on the concept of 'metasurface bit' [27], Tiejun Cui et al. introduced 'coding metasurface' and 'programmable metasurface' [28], making it possible to tune the electromagnetic properties of a single metasurface using electric signals. By introducing advanced algorithms, novel applications can be achieved with programmable metasurfaces, such as programmable hologram [29], programmable imager [30], and self-adapted clock [31]. However, it requires external voltage signals to get the coding metasurface tunable, and thus power supply and control circuits are essential to provide DC bias. Our work focuses on achieving a reconfigurable metasurface without applying any kind of signal. Our exploration also reveals the possibility of physically combining several centimeter-scale metasurface pieces into large-size and multi-functional metasurfaces.

In this article, we introduce the concept of metasurface tessellation, which involves tessellating metasurface tiles. A meta-tile consists of several types of unit cells with different physical parameters. We designed three specific types of meta-tiles, of dimensions 4×4, 6×6, and 9×9, using basic 0/1 coding unit cells. Then, with numerical simulation results, we would demonstrate how tessellations with different orientation distributions of meta-tiles perform different far-field scattering characteristics. In order to design the tessellation efficiently, we utilize a genetic algorithm to search for the best orientation distribution of the tiles for the exact far-field pattern we desire. The achieved far-field scattering patterns include reflecting incident waves to two symmetrically oriented directions with flexible elevation and azimuth angles, or four symmetrically oriented directions, or without advantageous orientation, similar to diffuse reflection.

## 2. Results

We start with the construction of a metasurface tile, which is composed of many elements; but is far smaller than the entire metasurface. To comprise the tile, each of the elements may possess their electromagnetic properties (transmission, reflection, phase, amplitude, etc.) at a specific frequency. The arrangement of elements on a tile could either be arbitrary or specially designed for particular functions. A regular $n$-sided meta-tile will remain its initial profile when rotated by $2\pi/n$ or any integer multiple of this angle. However, interference performance among different meta-tiles could vary significantly. In other words, the meta-tiles in different orientations can be tessellated into different metasurface patterns, with the scattering performance adjusted by the orientation distribution of tiles rather than the phase tuning of each unit.

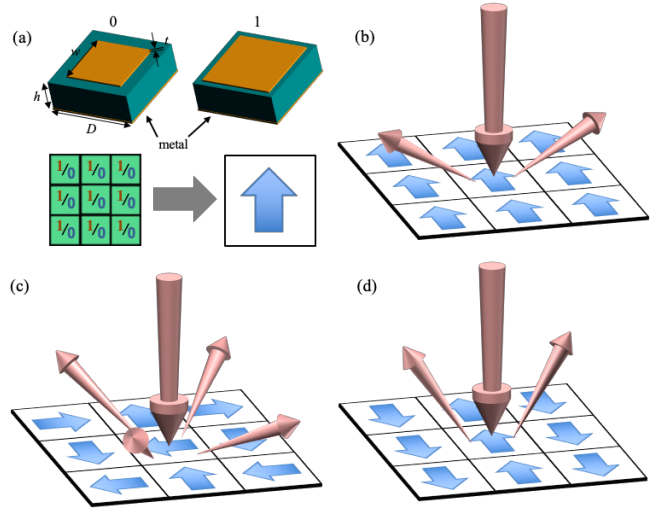

Fig. 1. Schematic illustration of metasurface tiles and orientation tessellations. (a) 0/1 coding elements and the construction of a coding-elements-based tile. (b)-(d) Orientation tessellations which scatter waves, separately, to different directions. (b), and (d) Metasurface tessellations to reflect the incident beam to two symmetric directions of different azimuth angles. (c) Another tessellation to reflect the incident beam to four symmetric directions.

To illustrate the above assumption, we utilize the concept of coding metasurface, where '0' and '1' represent two kinds of elements with a phase difference of $\pi$. If we define the '0' elements with a '0' phase response, then the '1' elements will have a $\pi$ phase response.

Therefore, we can characterize the phase property of a metasurface with a coding matrix, where each element, either '0' or '1', represents the scattering phase of an element in the metasurface. Many structures have been proven to meet this requirement, such as a square metallic patch printed on a dielectric substrate that lies on top of a metallic board (Fig. 1 (a)). The substrate has a thickness of $h = 1.964$ mm with a dielectric constant of 2.65 and a loss tangent of 0.001. The patch and the metallic board underneath have a thickness of $t = 0.018$ mm, width of $w$, and the periodicity of the unit cell is $D = 5$ mm. When the patch widths are designed to be 4.8 and 3.75 mm, the phase difference is nearly 180° from 8.1 GHz to 12.7 GHz [28]. With the '0' and the '1' elements, a square meta-tile can be formed with $P \times P$ of the elements (Fig. 1 (a)). Then, $Q \times Q$ pieces of such tiles, each rotated to a given orientation, can combine into an oriented tessellation with specific scattering properties (Figs. 1(b–d)). Here, elements on a fixed tile constitute one of the submatrices of the coding matrix.

Considering the general situation where the elements are not rotationally symmetric, each meta-tile can rotate by 0°, 90°, 180°, or 270° to generate four submatrices in total. If we define the zero-rotated tile (or submatrix) as N (for North), then the 90°, 180°, and 270° counterclockwise rotated tiles are marked as W, S, and E (for West, South, and East), respectively. $Q \times Q$ of the tiles are tessellated into a metasurface with $(P \cdot Q) \times (P \cdot Q)$ elements, which is characterized by a coding matrix that consists of $Q \times Q$ submatrices. The orientation distribution of the tiles should follow a 'layout matrix' where each element, being either 'N', 'W', 'S', or 'E', also stands for the submatrix on its place. For example, Fig. 2 (a) demonstrates a specific meta-tile. On the left of Fig. 2 (b) is a layout matrix, following which the tiles are combined into a metasurface. The right part of Fig. 2 (b) shows the entire tessellation with the calculated far-field scattering pattern. We present three typical meta-tiles of dimensions 4×4, 6×6, and 9×9 (Figs. 2 (a), 3 (a), and 4 (a)), and exemplarily show how we design the orientation distribution of the tiles to obtain the desired scattering patterns. All the following calculations are based on the above element, and we let $\lambda/D = 5.8$, where $\lambda$ is the wavelength of the incident beam.

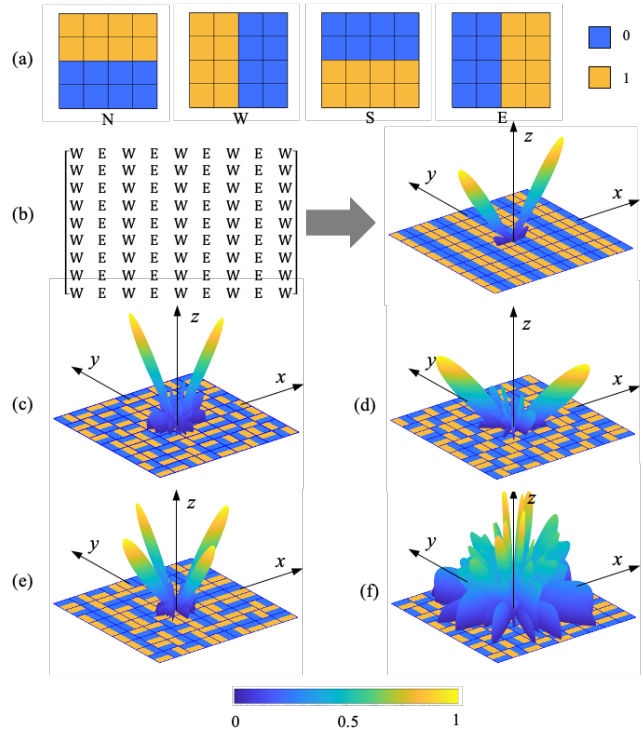

Fig. 2. A design of meta-tiles consisting of 4×4 coding elements and several tessellations with their far-field scatter patterns under the normal incident EM waves. (a) By rotating the tile in the *xy* plane, we obtain four coding submatrices, marked as N, W, S, and E, respectively. (b) The 9×9 tiles are combined according to a layout matrix. The tessellation can reflect the beam in two directions. (c-f) Tiles are tessellated by different layout matrices designed by optimization algorithms to perform different scattering patterns.

To begin with, consider meta-tiles of dimensions 4×4 (Fig. 2 (a)). They can be tessellated according to a layout matrix of dimensions 9×9, as shown in the left part of Fig. 2 (b), where W and E elements can be separately replaced by submatrices

$$W = \begin{bmatrix} 1 & 1 & 0 & 0 \\ 1 & 1 & 0 & 0 \\ 1 & 1 & 0 & 0 \\ 1 & 1 & 0 & 0 \end{bmatrix}, \text{ and } E = \begin{bmatrix} 0 & 0 & 1 & 1 \\ 0 & 0 & 1 & 1 \\ 0 & 0 & 1 & 1 \\ 0 & 0 & 1 & 1 \end{bmatrix} \quad (1)$$

to obtain a coding matrix of dimensions 36×36 consisting of '0' and '1' to stand for the phase response of each element in the metasurface. The scattering phase of the meta-element in row *m* and column *n* is assumed to be $\varphi(m, n)$, which is either 0 or $\pi$ depending on the element in the row *m*, and column *n* of the coding matrix. Hence, under a normal incident wave, the far-field function scattered by the metasurface is expressed as [28, 32]

$$F(\theta,\varphi) = \sum_{m=1}^{PQ} \sum_{n=1}^{PQ} \exp\left\{-i\left\{\varphi(m,n) + kD\cos\theta\left[(m-1/2)\cos\varphi + (n-1/2)\sin\varphi\right]\right\}\right\} \quad (2)$$

where $\theta$ and $\varphi$ are the elevation and azimuth angles of an arbitrary direction, respectively. The metasurface is a tessellation of $Q \times Q$ pieces of tiles, and each tile is a mixture of $P \times P$ meta-elements. The far-field function is calculated and normalized to be drawn on the right side of Fig. 2 (b).

Note that, with a phase difference of $\pi$, the '0' and '1' elements are switchable with each other. Therefore, it is mathematically derivable that $|F(\theta,\varphi)| = |F(\theta,\varphi + \pi)|$, which means the scattering pattern must be central symmetrical. Asymmetrical patterns can be achieved with tiles consisting of more than binary phase elements. In Fig. 2 (b), the incident wave was reflected in two symmetrical directions within the *xz* plane. We can tailor its scattering by designing the orientation distribution of the meta-tiles, and optimization algorithms can be applied, such as the genetic algorithm (GA) [22, 33, 34]. For 9×9 pieces of the tiles, we adopted the GA to optimize the orientation distribution to achieve several specific scattering patterns. In GA, the initial gene sequences are decoded into a group of layout matrices, and we set the objective function of GA to equal the far-field function in a specific direction $|F(\theta_0,\varphi_0)|$. Then, the algorithm helps to find an optimal layout matrix, according to which the tiles can be put together to reflect the beam in the set direction as well as its symmetrical one. For instance, the optimized layout matrices for ($\theta_0 = \pi/3$, $\varphi_0 = \pi/4$) and ($\theta_0 = \pi/6$, $\varphi_0 = \pi/4$) are

$$\begin{bmatrix} E & S & N & N & S & S & W & W & S \\ S & W & N & E & S & W & N & S & E \\ W & N & S & E & W & W & S & S & N \\ N & E & S & W & N & S & E & W & W \\ E & S & W & N & S & E & N & W & E \\ E & N & W & S & S & N & N & E & E \\ E & W & S & E & N & N & N & S & N \\ W & S & S & W & N & E & E & W & W \\ S & S & E & W & S & S & W & N & E \end{bmatrix} \text{ and } \begin{bmatrix} W & S & W & S & S & W & S & W & E \\ S & W & S & W & W & S & W & E & W \\ N & E & W & N & S & W & E & N & E \\ S & W & W & S & W & S & W & E & S \\ W & N & E & N & S & N & S & E & N \\ W & E & W & S & N & S & E & W & S \\ S & W & E & W & S & S & N & S & N \\ N & E & N & E & E & N & E & N & S \\ E & N & S & E & W & S & W & E & N \end{bmatrix}, \quad (1)$$

respectively. The integrated metasurfaces together with their scattering patterns, are depicted in Figs. 2 (c) and 2 (d), respectively. It is also possible for the metasurface tessellation to reflect the beam mainly to four orientations. We set the objective function to equal the minimum between $|F(\pi/4,0)|$ and $|F(\pi/4,\pi/2)|$ so that we can maximize the far-field function in the two directions simultaneously. The function values in directions that are centro-symmetric to our target are also optimized, so the scattering in four directions is enhanced, with the optimized layout matrix

$$\begin{bmatrix} S & E & S & E & W & E & W & E & S \\ N & E & N & N & N & N & W & E & N \\ W & E & S & E & S & E & W & E & S \\ W & E & W & E & N & E & N & N & N \\ W & E & W & S & S & S & S & S & S \\ W & N & N & E & W & E & W & N & N \\ W & S & W & S & W & E & S & S & W \\ W & N & W & E & N & N & N & E & N \\ S & E & S & E & W & E & S & S & S \end{bmatrix}, \quad (2)$$

as shown in Fig. 2 (e). Furthermore, we can also utilize GA to minimize the overall reflection, by setting the objective function of the GA to a negative maximum of $|F(\theta,\varphi)|$, $\theta$ ranging from 0 to $\pi/2$ and $\varphi$ from 0 to $\pi$. Then, the optimized layout matrix becomes

$$\begin{bmatrix} S & N & E & S & W & S & E & S & N \\ E & E & N & E & N & W & S & W & W \\ N & W & N & N & E & S & E & E & E \\ W & S & N & E & E & E & E & W & E \\ N & W & W & W & S & E & E & S & N \\ W & N & S & N & S & E & E & S & S \\ S & N & N & E & S & N & E & E & S \\ E & E & E & W & S & N & N & N & N \\ N & W & S & N & S & N & N & N & N \end{bmatrix}, \quad (3)$$

The far-field scattering pattern of the tessellated metasurface is shown in Fig. 2 (f), which is almost a diffuse reflection.

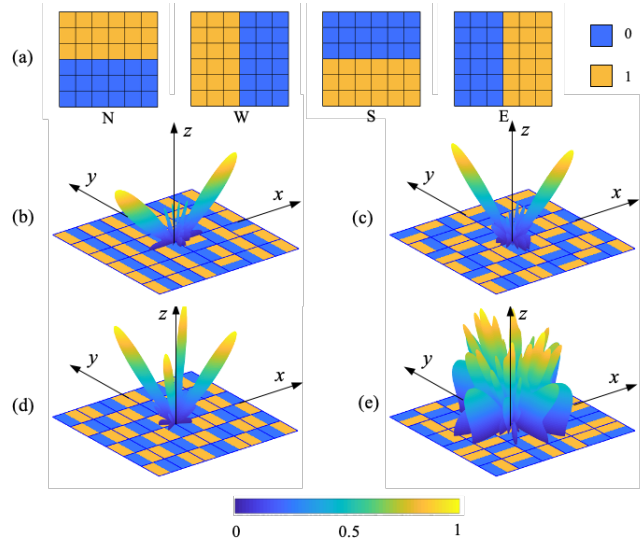

Fig. 3. A construction of meta-tiles of dimensions 6×6, and several tessellations for different scattering characteristics. (a) The 6×6 meta-tile in different orientations. (b-e) 6×6 pieces of the tiles are combined according to different orientation distributions to constitute metasurfaces with distinct scattering patterns, optimized by genetic algorithm, similar to that shown in Fig. 2.

For meta-tiles of dimensions 6×6, a similar process was followed. The designed tile is shown in Fig. 3 (a). By arranging 6×6 pieces of tiles into a GA-optimized layout, the tessellated metasurface can reflect waves to two symmetric orientations with a tunable elevation angle $\theta_0$, either within the $xz$ plane (Fig. 3 (b)) or away from the $xz$ plane with a specified azimuth angle $\varphi_0$ (Fig. 3(d)). The EM waves can also be reflected in four orientations, and diffuse reflection is also accessible (Figs. 3 (e) and 3 (f)). The orientation distribution matrices for Figs. 3 (b)-(e) are

$$\begin{bmatrix} W & W & W & E & E & E \\ W & W & W & E & N & E \\ W & W & W & N & E & E \\ W & W & W & N & E & S \\ W & W & W & E & E & E \\ W & W & W & N & E & E \end{bmatrix}, \begin{bmatrix} S & N & E & N & E & N \\ W & S & N & S & N & E \\ S & N & S & N & E & W \\ W & E & W & E & N & E \\ S & W & E & N & E & W \\ W & E & W & S & N & E \end{bmatrix},$$

$$\begin{bmatrix} E & W & E & W & E & W \\ W & E & W & E & W & E \\ E & W & E & W & E & W \\ W & E & W & E & W & E \\ E & W & E & W & E & W \\ W & E & W & E & W & E \end{bmatrix} \text{ and } \begin{bmatrix} S & N & S & S & E & S \\ E & S & E & S & W & S \\ E & W & W & S & W & E \\ S & S & N & W & N & N \\ N & S & E & E & S & S \\ E & N & E & E & N & N \end{bmatrix}, \quad (4)$$

respectively.

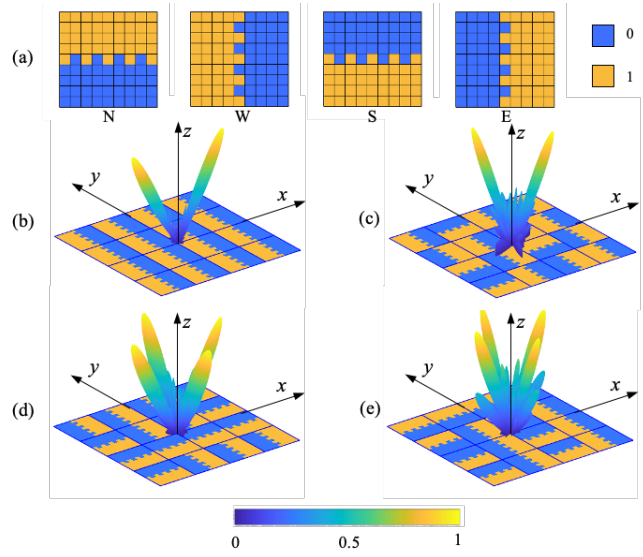

Fig. 4. (a) A type of meta tile of dimensions 9×9. (b, c) Tessellations of 4×4 pieces of tiles to maximize the scattering in tunable two directions, optimized by GA. (d) Achievement of enhanced scattering in four directions. (e) Failure to achieve an excepted scattering pattern to be enhanced in very specified directions, even with GA optimization.

It is noted that this method is only effective when the tessellation comprises a large number of meta-tiles, or in other words when the tiles are far smaller than the entire metasurface. For general tessellation, the scattering is reconfigurable in some degrees of freedom, but there is a limitation on the reconfigurability. However, not all scattering patterns can be achieved using a single kind of meta-tile, especially for larger tiles within a normal-sized metasurface. Here, we try to combine 4×4 pieces of 9×9 tiles, as represented in Fig. 4 (a), to perform reconfigurable scattering. The GA works well for some desired patterns with flexible elevation and azimuth angle. For example, we adapt the GA to maximize the reflection on ($\theta = \pi/4$, $\varphi = 0$) and obtain an optimized layout matrix

$$\begin{bmatrix} W & W & W & W \\ W & W & W & W \\ W & W & W & W \\ W & W & W & W \end{bmatrix}, \quad (5)$$

according to which the tessellated metasurface shows the desired scattering pattern in Fig. 4 (b). The enhanced scattering on ($\theta = \pi/3$, $\varphi = \pi/4$) is also achievable, as shown in Fig. 4 (c), with the optimized layout matrix

$$\begin{bmatrix} W & S & W & E \\ E & W & S & W \\ N & E & N & E \\ S & W & E & W \end{bmatrix}. \quad (6)$$

The enhanced scattering in four directions is also achievable. When we set the objective function to be the minimum between $|F(\pi/4,0)|$ and $|F(\pi/4,\pi/2)|$, the GA can find the best orientation distribution

$$\begin{bmatrix} E & E & E & S \\ S & E & E & S \\ S & S & S & S \\ S & E & E & E \end{bmatrix}, \quad (7)$$

for the meta-tiles to reflect the beams into four directions, as shown in Fig. 4 (d). However, when we try to enhance the scattering on ($\theta_0 = \pi/3$, $\varphi_0 = 0$) and ($\theta_1 = \pi/3$, $\varphi_1 = \pi/2$) as well as their symmetric directions simultaneously, the algorithm fails to find a flawless orientation distribution for the tiles to arrange into the desired scattering pattern. This is because a layout matrix of $Q \times Q$ dimensions has only $4^{Q \times Q}$ alternatives, and the fewer pieces we adopt, or the larger meta-tile we adopt for a metasurface of given dimensions, the flexibility worsens.

Our design results are based on the three typical kinds of meta-tiles with binary elements, and every tessellation comprises only one type of meta-tile. In order to achieve more flexibility in reconfiguring the scattering pattern, each meta-tile can be comprised of more than two kinds of phase elements. On the other hand, more types of meta-tiles with distinct element distributions can be integrated into a tessellation. The tiles can be tessellated even in an unaligned way, which enables much more flexibility into the metasurface tessellation design. The modified methods are demonstrated in Fig. 5.

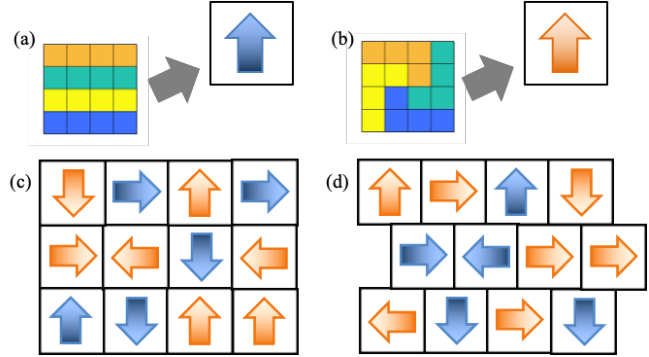

Fig. 5. Modified methods to increase the reconfigurability of the tessellated metasurface. (a) A construction of meta-tiles comprising 4 kinds of phase elements. (b) Another type of meta-tiles of the same dimensions, with element distribution distinct to (a). (c) Different tiles arranged together to make up a tessellation. (d) Tiles are arranged into an unaligned tessellation for better reconfigurability.

## 3. Conclusion

In this letter, we have proposed the concept of arranging small meta-tiles into metasurface tessellation with reconfigurable scattering patterns and also present three typical kinds of meta-tiles consisting of binary meta-elements. It is possible to tailor the construction of tiles as well as their orientation distribution to tune the scattering pattern of the tessellation. We utilized a genetic algorithm to determine the optimal orientation distribution of meta-tiles to perform diverse scattering patterns. Our work can be extended to tessellation consisting of tiles that are of various shapes, such as polygons or circles, and which consist of multiple kinds of meta-elements.

## Funding

## Disclosures

The authors declare no conflicts of interest.